%% file: arxiv.tex
\documentclass{emulateapj}
\usepackage{url}
\submitted{ApJ, in press}

\newcommand{\axp}{1E~1547.0$-$5408}
\input{def}

\begin{document}
\title{\emph{CHANDRA} and \emph{RXTE} Observations of \axp:
Comparing the 2008 and 2009 Outbursts}
\shorttitle{\emph{Chandra} and \emph{RXTE} Observations of 1E 1547.0$-$5408}
\shortauthors{Ng et al.}
\author{C.-Y.~Ng\altaffilmark{1}, V.~M.~Kaspi, R.~Dib, S.~A.~Olausen, P.~Scholz}
\affil{Department of Physics, McGill University, Montreal, QC H3A 2T8, Canada}
\email{ncy@hep.physics.mcgill.ca}
\altaffiltext{1}{Tomlinson Postdoctoral Fellow}
\author{T.~G\"{u}ver, F.~\"{O}zel}
\affil{Department of Astronomy, University of Arizona, Tucson, AZ, 85721}
\author{F.~P.~Gavriil}
\affil{NASA Goddard Space Flight Center, Astrophysics Science Division, Code 662, Greenbelt, MD 20771}
\affil{Center for Research and Exploration in Space Science and Technology, University of Maryland Baltimore County, Baltimore, MD 21250}
\and
\author{P.~M.~Woods}
\affil{Corvid Technologies, Huntsville, AL 35806}

\begin{abstract}
We present results from \emph{Chandra X-ray Observatory} and
\emph{Rossi X-ray Timing Explorer} (\emph{RXTE}) observations
of the magnetar 1E 1547.0$-$5408 (SGR J1550$-$5418) following the
source's outbursts in 2008 October and 2009 January. During the time span of
the \cxo\ observations, which covers days 4 through 23 and days 2 through
16 after the 2008 and 2009 events, respectively, the source spectral
shape remained stable in the \cxo\ band, while the pulsar's spin-down
rate in the same span in 2008 increased by a factor of 2.2 as measured
by \emph{RXTE}. This suggests decoupling between the source's spin-down and
radiative changes, hence between the spin-down-inferred magnetic field
strength and that inferred spectrally. The lack of spectral variation during flux decay is surprising for models of magnetar outbursts.
We also found a strong
anti-correlation between the phase-averaged flux and the pulsed
fraction in the 2008 and 2009 \cxo\ data, but not in the pre-2008
measurements.  We discuss these results in the context of the magnetar model.
\end{abstract}


\section{Introduction} 
Anomalous X-ray pulsars (AXPs) and soft gamma repeaters (SGRs),
though previously thought to be different classes of objects,
are now believed to all be strongly magnetized neutron stars,
known as ``magnetars''. They are characterized by long spin periods
of 2-12\,s and large spin-down rates that imply ultra-strong surface
dipole magnetic fields of $10^{14}$-$10^{15}$\,G (see reviews
by \citealt{kas07} and \citealt{mer08}). The most remarkable feature
of magnetars is their violent outbursts, during which the X-ray
luminosity could increase by a few orders of magnitude. In
the context of the twisted magnetosphere model \citep{tlk02}, the
energy release is due to magnetic field re-arrangement, which
is possibly triggered by crustal deformation. While the
post-outburst behavior of a magnetar can provide important
information on the physical conditions of the magnetosphere
and the stellar surface, only a handful of follow-up studies have 
previously been carried out with focusing X-ray instruments
\citep[e.g.][]{wkt+04,gkw06,icd+07}, because the transient nature
of these events requires prompt observations. In this work, we study
recent outbursts from the AXP \object{\axp}\footnote{Also known
as SGR J1550$-$5418 or PSR J1550$-$5418.} in 2008 and 2009 using
observations made with the \emph{Chandra X-ray Observatory} and
the \emph{Rossi X-ray Timing Explorer} (\emph{RXTE}).

\begin{deluxetable*}{lccccccc}
\tablecaption{{\it Chandra} Observation Parameters and Results
\label{t1}}
\tablewidth{0pt}
\tabletypesize{\scriptsize}
\tablehead{\colhead{Obs.} & \colhead{Date} & \colhead{MJD} &\colhead{ObsID} & \colhead{Days Since} &
\colhead{Exposure} & \colhead{Count Rate\tablenotemark{a}} &
\colhead{Pulsed Fraction\tablenotemark{a}}\\
& & & & \colhead{Outburst} & \colhead{(ks)} & \colhead{(s$^{-1}$)} }
\startdata
\multicolumn{8}{c}{2008 October}\\\hline
1 & 2008 Oct 7 & 54746.6 & \dataset[ADS/Sa.CXO#obs/8811]{8811}
& 4.2 & 12.1 & 1.35(1) & 0.21(1)\\
2 & 2008 Oct 10 & 54749.5 & \dataset[ADS/Sa.CXO#obs/8812]{8812}
& 7.2 & 15.1 & 1.17(1) & 0.22(1)\\
3 & 2008 Oct 18 & 54757.6 & \dataset[ADS/Sa.CXO#obs/8813]{8813}
& 15.2 & 10.1 & 1.15(1) & 0.33(1)\\
4 & 2008 Oct 21 & 54760.8 & \dataset[ADS/Sa.CXO#obs/10792]{10792}
& 18.4 & 10.1 & 1.07(1) & 0.35(1)\\
5 & 2008 Oct 26 & 54765.1 & \dataset[ADS/Sa.CXO#obs/8814]{8814}
& 22.8 & 23.0 & 0.99(1) & 0.31(1)\\\hline
\multicolumn{8}{c}{2009 January}\\\hline
6 & 2009 Jan 23 & 54855.0 & \dataset[ADS/Sa.CXO#obs/10185]{10185}
& 2.0 & 10.1 & 0.95(1)\tablenotemark{b} & 0.09(1)\\
7 & 2009 Jan 25 & 54856.7 & \dataset[ADS/Sa.CXO#obs/10186]{10186} & 3.7 & 12.1 & 3.1(2) & 0.09(1)\\
8 & 2009 Jan 29 & 54860.8 & \dataset[ADS/Sa.CXO#obs/10187]{10187} & 7.8 & 13.1 & 2.5(1) & 0.13(1)\\
9 & 2009 Feb 06 & 54868.6 & \dataset[ADS/Sa.CXO#obs/10188]{10188} & 15.6 & 14.3 & 2.2(1) & 0.12(1)
\enddata

\tablenotetext{a}{In the $1-7$\,keV energy range.}
\tablenotetext{b}{Obs.\ 6 was made with the HETG, which results in a reduced count rate.
Correcting for the effective area gives a count rate of $\sim3.6$.}
\end{deluxetable*}

The X-ray source \axp\ was discovered by \citet{lm81} with the \emph{Einstein
Observatory}. Based on the X-ray spectrum and infrared flux,
\citet{gg07} first suggested the source as a magnetar candidate.
The detection of radio pulsations by \citet{crh+07} directly
confirmed the pulsar nature of this object; its spin period
of 2.1\,s is shorter than any other known magnetars.\footnote{See
the online magnetar catalog at
\url{http://www.physics.mcgill.ca/~pulsar/magnetar/main.html}.}
The pulsar's spin-down rate as reported by \citet{crh+07} implies
a surface magnetic dipole field of $2.2\times10^{14}$\,G.
\citet{hgr+08} reported a high state of this magnetar in 2007 based
on \xmm\ observations, and they concluded that an X-ray outburst
had occurred between 2006 and 2007. On 2008 October 3
(MJD 54742), \axp\ showed bursting activity with outbursts
detected by \emph{Swift} \citep{ier+10} and by the Gamma-ray Burst
Monitor (GBM) onboard the \emph{Fermi Gamma-ray Space Telescope}
\citep{kgk+10}. On 2009 January 22 (MJD 54853), the AXP entered a
second active phase. More than 200 bursts were detected within
a few hours by \emph{Swift}, \emph{International Gamma-Ray
Astrophysics Laboratory} (\emph{INTEGRAL}), \emph{Fermi} GBM,
and \emph{Suzaku} WAM \citep{ghm+09,mgw+09,snb+10,kgk+10,ttu+09}.
Follow-up imaging observations with \emph{Swift} and \xmm\ revealed
dust scattering-X-ray rings centered on the source, from which
\citet{tve+10} deduced a source distance of 3.9\,kpc. Based on
\emph{Suzaku} observations taken 7\,days after the 2009 outburst,
\citet{enm+10} reported a hard power-law tail in the spectrum, of
photon index $\Gamma$ between 1.33 and 1.55, and extending up to at least
110\,keV. Hard X-ray pulsations were also detected by \emph{INTEGRAL} 
in the 20-150\,keV band \citep{kdh09,dkh09}. The pulsed emission
has $\Gamma=1.55$ and the spectral shape remained stable over the
observation period covering from day 2 to day 9 after the outburst.

In this paper, we report on results from \cxo\ and \emph{RXTE}
observations taken after the 2008 and 2009 events. The observations
and results are reported in Section \ref{s2}, and we discuss
their physical implications in Section \ref{s3}. We summarize our
findings in Section \ref{s4}.

\begin{figure*}[ht]
\epsscale{0.8}
\plotone{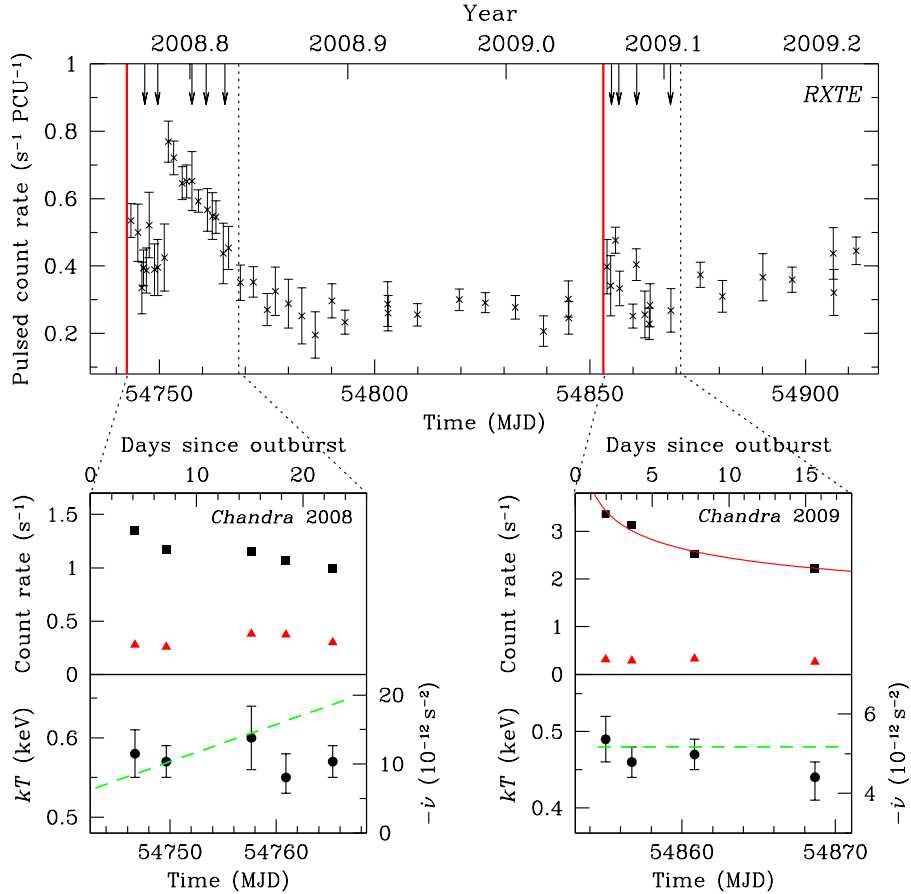}
\caption{Top: rms pulsed flux of \axp\ in the 2-10\,keV
range obtained with \emph{RXTE}. The red vertical solid lines
mark the outbursts in 2008 and 2009, and the arrows indicate the
\cxo\ observation epochs. Bottom: total and rms pulsed
\cxo\ count rates in the 1-7\,keV band, shown by black squares
and red triangles in the upper panels, respectively. The
statistical uncertainties are negligible. The solid line in 2009
illustrates a power-law fit to the flux decay. Note that the
first \cxo\ observation in 2009 was made with the HETG, which
precludes a direct comparison of the count rates. Therefore, the
data point plotted here is estimated from the spectral analysis
results. The lower panels show the best-fit blackbody temperature
from the PL+BB model, as listed in Table~\ref{t2}. The green
dashed lines indicate the spin-down rate obtained from the phase-coherent
\emph{RXTE} timing solutions. Uncertainties in $\dot\nu$ measurements
are negligible (at 1\% level).
\label{f1}}
\end{figure*}

\section{Observations and Results}
\label{s2} 
\subsection{{\it RXTE} Observations and Results} 
Since the 2008 October outburst, \axp\ has been monitored regularly
with \emph{RXTE}. Data are collected with the Proportional
Counter Array (PCA) instrument that consists of five collimated
xenon/methane multi-anode Proportional Counter Units (PCUs).
Only \texttt{GoodXenonwithPropane} mode data were used in our
analysis, which give 1\,$\mu$s time resolution and 256 energy channels
in the 2-60\,keV energy range. We considered events from only
the top Xenon layer of each PCU to maximize the signal-to-noise
ratio (S/N). Typical exposure times were in the range 2--7~ks.
In all, we report here on a total of 55 observations taken 
between MJDs 54743 and 54911. A more complete report on the
{\it RXTE} data will be presented elsewhere (R.\ Dib et al.,
in preparation).

\subsubsection{{\it RXTE} Timing}
\label{sec:rxtetiming} 
To determine the pulsar spin parameters, we cleaned the \emph{RXTE}
data and selected events between 2 and 6.5\,keV. As the source exhibited
many short X-ray bursts, we removed all burst intervals for this
analysis. The photon arrival times were first corrected to the solar
system barycenter, then binned with 31.25\,ms time resolution. The time
series were folded at the nominal spin period and pulse arrival times
were extracted by cross-correlating with a template. These arrival
times were then fitted to a simple phase-coherent timing model using
the TEMPO software package.\footnote{\url{http://www.atnf.csiro.au/research/pulsar/tempo/}} 
Details on the phase folding and periodicity
search techniques are described in \citet{dkg09}. Our best-fit timing
solution gives a spin frequency $\nu=0.48277818(5)$\,Hz, frequency
derivative $\dot\nu=-6.6(1)\times10^{-12}$\,s$^{-2}$, and second
frequency derivative $\ddot\nu=-6.2(1)\times10^{-18}$\,s$^{-3}$
covering the 2008 \cxo\ epochs (reference epoch of MJD 54743.0), and
$\nu=0.4825962(4)$\,Hz, $\dot\nu=-5.17(5)\times10^{-12}$\,s$^{-2}$
with negligible $\ddot\nu$ for the 2009 {\it Chandra} epochs (reference epoch of MJD 54854.0).
These values are consistent with those reported by
\citet{ier+10} and \citet{kgk+10} based on the 2008 \emph{Swift}
and 2009 \emph{Fermi} observations, respectively.

\subsubsection{{\it RXTE} Pulsed Fluxes}
\label{s232}
Once the timing solution was obtained, we re-generated the pulse profiles
in the 2-10\,keV band, and calculated the rms pulsed flux according
to the formula in \citet{dkg08}, using seven harmonics. The results
are plotted in Figure~\ref{f1} and reveal a complicated flux
evolution. In this paper, we focus mainly on time periods near
the \cxo\ epochs, and a full analysis of the \emph{RXTE} data will
be presented by R.\ Dib et al.\ (in preparation). Approximately 11 days after
the initial 2008 October 3 (MJD 54742) trigger, during which the pulsed
flux decayed roughly monotonically, the pulsed emission abruptly
increased again by 80\% between two \emph{RXTE} observations taken
on MJDs~54751.2 and 54752.1, then decayed monotonically again until it
reached a minimum level on MJD~54786.2. As the onset of this second
event is not resolved, we report an upper limit of 1\,day for
the rise time. This second flux enhancement decayed monotonically
for $\sim$34 days; the pulsed count rate in this period can be
parameterized by an exponential fall-off with $1/e$ decay time of
$\sim25$ days, although a linear decay is also consistent with the data.
Interestingly, the 2009 event, which is more energetic, is far less
dramatic as seen by {\it RXTE}, with a relatively small increase in the pulsed
flux observed.

\subsection{\cxo\ Observations and Results}
The 2008 outburst of \axp\ triggered a series of \cxo\ 
observations through our Target of Opportunity (ToO) program.
Five total pointings were made on days 4, 7, 15, 18 and 23 after
the outburst, with exposures ranging from 12 to 23\,ks. Data
were taken with the ACIS-S detector in continuous clocking (CC)
mode, which has a time resolution of 2.85\,ms. For the 2009 event,
another \cxo\ ToO program followed the outburst, with four
ACIS-S CC mode observations taken on days 2, 4, 8 and 16. The
first exposure in 2009 was taken with the High Energy Transmission
Grating (HETG). We include only the zeroth-order events in this
latter data set in our analysis. A summary of observation
parameters is provided in Table~\ref{t1}.

We performed all the \cxo\ data reduction using CIAO~4.2
with CALDB 4.2.0. We first removed the burst intervals 
in the data. Source counts were extracted from a
3\arcsec\ diameter aperture and background counts were from the
whole chip excluding the central 40\arcsec\ region (i.e., a total
width of 7\farcm7). We note that although there is a nearby
source XMMU J155053.7$-$541925 southwest to the pulsar
\citep{gg07}, the projected separation is always larger than
30\arcsec\ in the CC-mode data. Therefore, it does not
contaminate the source counts. The source count rates in the
1-7\,keV energy range are reported in Table~\ref{t1} and plotted
in Figure~\ref{f1}, in which the backgrounds have been accounted
for, although they are less than 0.5\%. The count rates
are well below the telemetry limit of \cxo, and pileup is
negligible due to the short frame time of the CC-mode exposures.
While the source flux changes between epochs, we found no
short-term variability within any individual exposure. Employing
a test algorithm suggested by \citet{gl92}, we found $<10\%$
probability of source variability within individual observations, which
is not statistically significant. The flux decay in 2009 can be modeled
by a power law of index $\alpha=-0.21\pm0.01$, which is plotted in
Figure~\ref{t1}. However, such a simple relation is not observed
in the 2008 flux evolution. In particular, the count rates are very
similar between the second and third exposures, which is likely related
to the enhanced pulsed flux detected by \emph{RXTE}.

\begin{figure}[th]
\epsscale{1.2}
\plotone{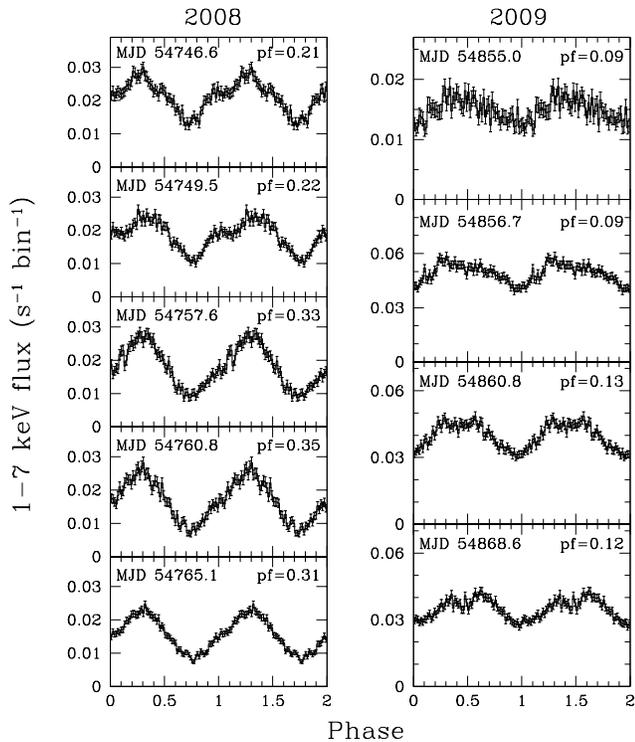}
\caption{Pulse profiles of \axp\ in 1-7\,keV obtained from the
\cxo\ observations, using 64 phase bins. The rms PFs
from Table~\ref{t1} are indicated.\label{f2}}
\end{figure}

\begin{figure}[th]
\begin{center}
\includegraphics[angle=270,width=0.4\textwidth]{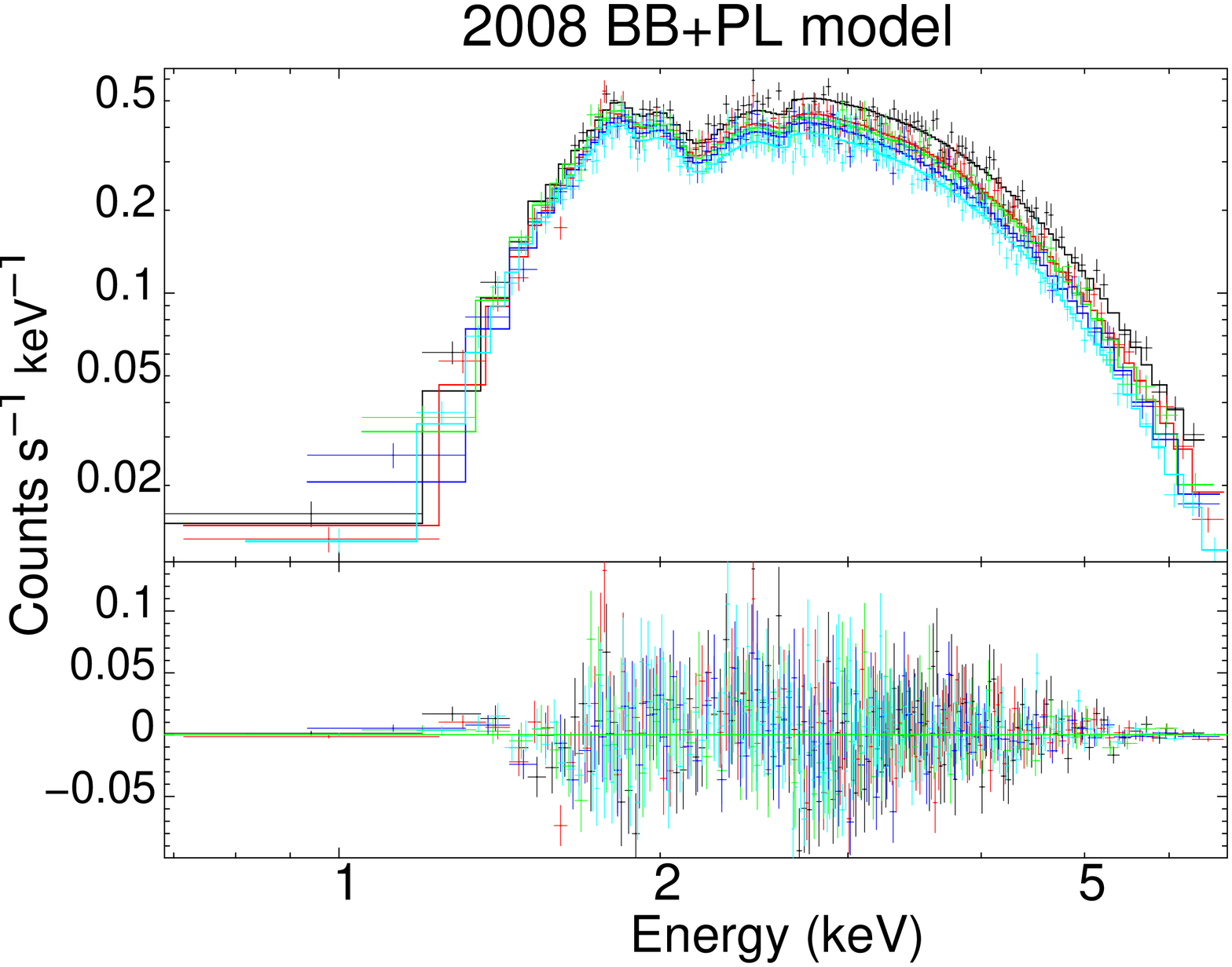}
\includegraphics[angle=270,width=0.4\textwidth]{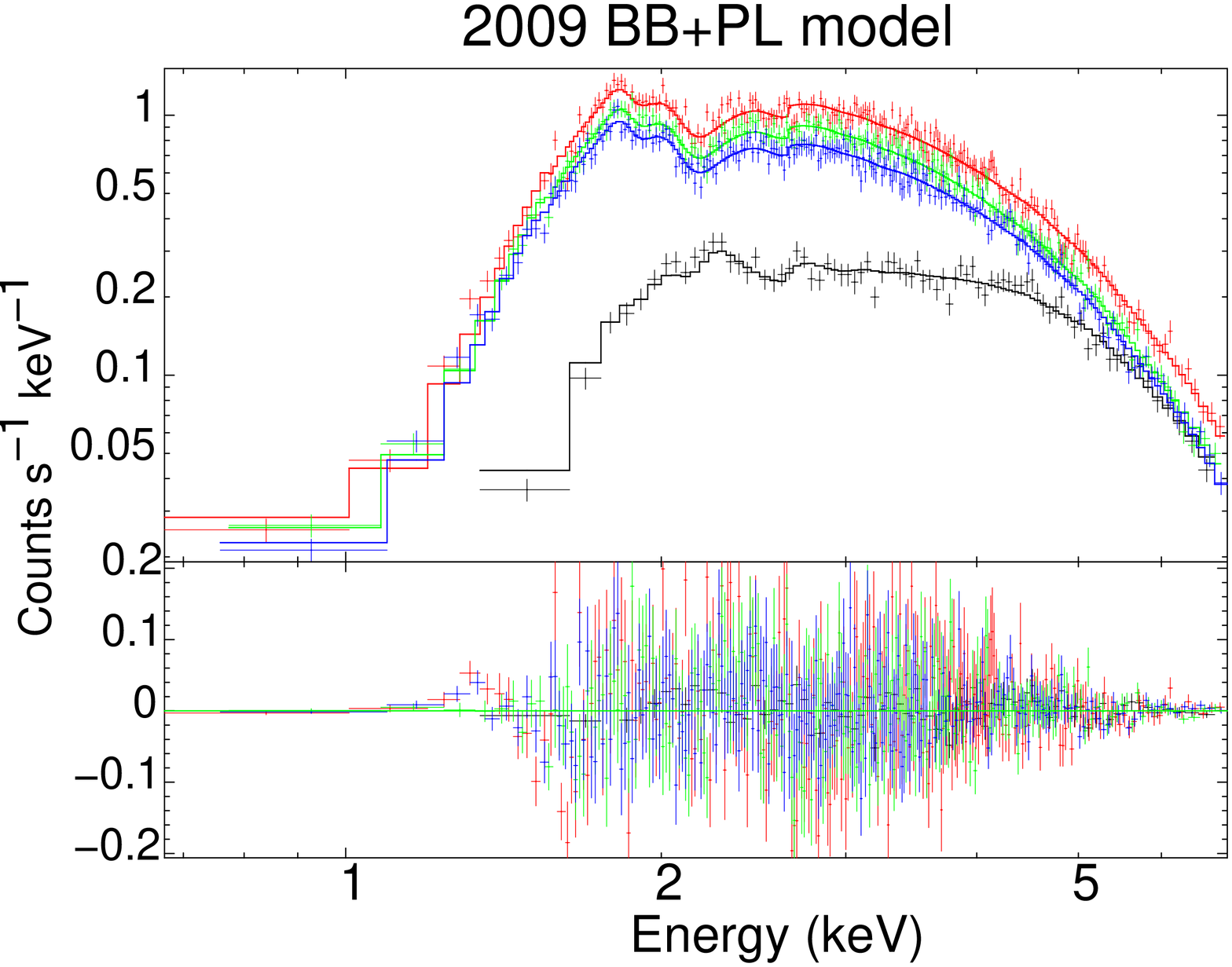}
\end{center}
\caption{Best-fit blackbody plus power-law model fit to the 2008
and 2009 \emph{Chandra} spectra (upper panels) with the corresponding
residuals (lower panels). Different observations are shown by different
colors. The corresponding best-fit spectral parameters are listed
in Table~\ref{t2}.
\label{f3}}
\end{figure}

\subsubsection{{\it Chandra} Timing}
We applied a barycenter correction to the \cxo\ data and then
folded the photon arrival times according to the \emph{RXTE} ephemeris.
The resulting lightcurves are shown in Figure~\ref{f2}. The pulse
profiles in 2008 suggest some hints of a multi-peak morphology in the
first two observations, then evolve into a single peak. By contrast,
the 2009 profiles exhibit a broad peak at first, with a second
peak emerging by the third observation. We found no obvious energy
dependence of the pulse shape across the \cxo\ band (0.5-10\,keV).
A direct comparison between the profiles in 2008 and 2009
indicates a much higher pulse modulation in the former. We
estimated the rms pulsed fractions (PFs) in 1-7\,keV, and the
results are listed in Table~\ref{t1}. We observe a clear trend
of increasing pulse modulation as the source recovers after the
outbursts.

\begin{deluxetable*}{lcccccccc}
\tablecaption{Phase-averaged Spectral Parameters of \axp\ for 
Different Models\label{t2}}
\tablewidth{0pt}
\tabletypesize{\scriptsize}
\tablehead{
\colhead{Obs.} & \colhead{$N_{\rm H}$} & \colhead{$B$} &
\colhead{$kT$} & \colhead{$R$} & 
\colhead{$\Gamma$} &
\colhead{$f^{\rm abs}$ \tablenotemark{a}} &
\colhead{$F_{\rm pl} /F_{\rm th} $ \tablenotemark{b}} &
\colhead{$\chi^2_\nu/\nu$}\\
& \colhead{($10^{22}$\,cm$^{-2}$)} &
\colhead{($10^{14}$\,G)} & \colhead{(keV)} & \colhead{(km)} &
& \colhead{($10^{-11}$\,erg} \\
& & & & & & \colhead{s$^{-1}$\,cm$^{-2}$)}
}
\startdata
\multicolumn{9}{c}{Blackbody + power-law model}\\\hline
2008\\
1& $4.1\pm0.1$\tablenotemark{c} & \nodata & $0.58\pm0.03$ & $2.0\pm0.3$ & $2.3_{-0.4}^{+0.3}$ & $1.92\pm0.03$ & $1.4_{-0.6}^{+1.0}$ & 1.08/1342\tablenotemark{c} \\
2& $4.1\pm0.1$\tablenotemark{c} & \nodata & $0.57\pm0.02$ & $2.1_{-0.2}^{+0.3}$ & $2.3_{-0.4}^{+0.3}$ & $1.64\pm0.02$ & $1.1_{-0.5}^{+0.8}$ & 1.08/1342\tablenotemark{c} \\
3& $4.1\pm0.1$\tablenotemark{c} & \nodata & $0.60\pm0.04$ & $1.7_{-0.3}^{+0.4}$ & $2.8\pm0.3$ & $1.57\pm0.03$ & $2.6_{-1.1}^{+1.6}$ & 1.08/1342\tablenotemark{c} \\
4& $4.1\pm0.1$\tablenotemark{c} & \nodata & $0.55^{+0.03}_{-0.02}$ & $2.2_{-0.3}^{+0.4}$ & $2.4_{-0.5}^{+0.3}$ & $1.55\pm0.03$ & $1.2_{-0.6}^{+1.1}$ & 1.08/1342\tablenotemark{c} \\
5& $4.1\pm0.1$\tablenotemark{c} & \nodata & $0.57\pm0.02$ & $1.8\pm0.2$ & $2.8_{-0.3}^{+0.2}$ & $1.34\pm0.02$ & $2.1_{-0.8}^{+1.2}$ & 1.08/1342\tablenotemark{c} \\
2009\\
6& $4.1\pm0.1$\tablenotemark{c} & \nodata & $0.49\pm0.03$ & $4.4_{-0.7}^{+0.8}$ & $2.0_{-0.4}^{+0.3}$ & $5.11\pm0.09$ & $1.6_{-0.8}^{+1.8}$ & 1.08/1342\tablenotemark{c} \\
7& $4.1\pm0.1$\tablenotemark{c} & \nodata & $0.46\pm0.02$ & $4.7_{-0.5}^{+0.7}$ & $2.1\pm0.2$ & $4.61\pm0.05$ & $1.8_{-0.6}^{+1.0}$ & 1.08/1342\tablenotemark{c} \\
8& $4.1\pm0.1$\tablenotemark{c} & \nodata & $0.47\pm0.02$ & $4.0_{-0.5}^{+0.6}$ & $2.3\pm0.2$ & $3.67\pm0.04$ & $2.2_{-0.8}^{+1.3}$ & 1.08/1342\tablenotemark{c} \\
9& $4.1\pm0.1$\tablenotemark{c} & \nodata & $0.44^{+0.02}_{-0.03}$ & $4.1_{-0.6}^{+0.8}$ & $2.4\pm0.2$ & $3.22\pm0.04$ & $2.7_{-1.0}^{+1.8}$ & 1.08/1342\tablenotemark{c} \\\hline
\multicolumn{9}{c}{RCS + hard power-law model}\\\hline
2008\\
1& $4.0\pm0.1$\tablenotemark{c} & \nodata & $0.47^{+0.05}_{-0.04}$ & \nodata & \nodata & $1.91\pm0.03$ & \nodata &1.09/1338\tablenotemark{c} \\
2& $4.0\pm0.1$\tablenotemark{c} & \nodata & $0.47\pm0.04$ & \nodata & \nodata & $1.63\pm0.02$ & \nodata &1.09/1338\tablenotemark{c} \\
3& $4.0\pm0.1$\tablenotemark{c} & \nodata & $0.38^{+0.07}_{-0.05}$ & \nodata & \nodata & $1.56\pm0.03$ & \nodata &1.09/1338\tablenotemark{c} \\
4& $4.0\pm0.1$\tablenotemark{c} & \nodata & $0.43^{+0.07}_{-0.03}$ & \nodata & \nodata & $1.53\pm0.03$ & \nodata &1.09/1338\tablenotemark{c} \\
5& $4.0\pm0.1$\tablenotemark{c} & \nodata & $0.41\pm0.04$ & \nodata & \nodata & $1.33\pm0.02$ & \nodata &1.09/1338\tablenotemark{c} \\
2009\\
6& $4.0\pm0.1$\tablenotemark{c} & \nodata & $0.32^{+0.05}_{-0.13}$ & \nodata & 1.33\tablenotemark{d} & $5.11\pm0.09$ & $0.37\pm0.08$ & 1.09/1338\tablenotemark{c} \\
7& $4.0\pm0.1$\tablenotemark{c} & \nodata & $0.17^{+0.16}_{-0.04}$ & \nodata & 1.33\tablenotemark{d} & $4.62\pm0.05$ & $0.37_{-0.12}^{+0.14}$ & 1.09/1338\tablenotemark{c} \\
8& $4.0\pm0.1$\tablenotemark{c} & \nodata & $0.32^{+0.04}_{-0.16}$ & \nodata & 1.33\tablenotemark{d} & $3.67\pm0.03$ & $0.30_{-0.09}^{+0.05}$ & 1.09/1338\tablenotemark{c} \\
9& $4.0\pm0.1$\tablenotemark{c} & \nodata & $0.13^{+0.18}_{-0.01}$ & \nodata & 1.33\tablenotemark{d} & $3.23\pm0.04$ & $0.27_{-0.08}^{+0.11}$ & 1.09/1338\tablenotemark{c} \\\hline
\multicolumn{9}{c}{STEMS + hard power-law model}\\\hline
2008\\
1& $4.5\pm0.1$\tablenotemark{c} & $2.61_{-0.08}^{+0.10}$ & $0.313_{-0.008}^{+0.011}$ & 10\tablenotemark{d} & \nodata & $1.90\pm0.03$ & \nodata &1.08/1329\tablenotemark{c} \\
2& $4.5\pm0.1$\tablenotemark{c} & $2.69_{-0.07}^{+0.10}$ & $0.314_{-0.005}^{+0.011}$ & 10\tablenotemark{d} & \nodata & $1.62\pm0.02$ & \nodata &1.08/1330\tablenotemark{c} \\
3& $4.5\pm0.1$\tablenotemark{c} & $2.57\pm0.06$ & $0.28_{-0.06}^{+0.02}$ & 10\tablenotemark{d} & \nodata & $1.56\pm0.03$ & \nodata &1.08/1331\tablenotemark{c} \\
4& $4.5\pm0.1$\tablenotemark{c} & $3.4_{-0.3}^{+0.2}$ & $0.35_{-0.02}^{+0.05}$ & 10\tablenotemark{d} & \nodata & $1.53\pm0.03$ & \nodata &1.08/1332\tablenotemark{c} \\
5& $4.5\pm0.1$\tablenotemark{c} & $2.62_{-0.04}^{+0.06}$ & $0.30_{-0.07}^{+0.005}$ & 10\tablenotemark{d} & \nodata & $1.33\pm0.02$ & \nodata &1.08/1333\tablenotemark{c} \\
2009\\
6& $4.5\pm0.1$\tablenotemark{c} & $5.4_{-1.4}^{+0.9}$ & $0.46_{-0.13}^{+0.09}$ & 10\tablenotemark{\dag} & 1.33\tablenotemark{\dag} & $5.08\pm0.09$ & $0.21_{-0.11}^{+0.09}$ & 1.08/1334\tablenotemark{c} \\
7& $4.5\pm0.1$\tablenotemark{c} & $3.01_{-0.09}^{+0.12}$ & $0.306\pm0.003$ & 10\tablenotemark{\dag} & 1.33\tablenotemark{\dag} & $4.60\pm0.05$ & $0.16\pm0.02$ & 1.08/1335\tablenotemark{c} \\
8& $4.5\pm0.1$\tablenotemark{c} & $3.6_{-0.1}^{+0.2}$ & $0.312_{-0.002}^{+0.012}$ & 10\tablenotemark{\dag} & 1.33\tablenotemark{\dag} & $3.66\pm0.04$ & $0.12_{-0.02}^{+0.03}$ & 1.08/1336\tablenotemark{c} \\
9& $4.5\pm0.1$\tablenotemark{c} & $3.0_{-0.1}^{+0.6}$ & $0.30_{-0.011}^{+0.02}$ & 10\tablenotemark{\dag} & 1.33\tablenotemark{\dag} & $3.23\pm0.04$ & $0.15_{-0.01}^{+0.02}$ & 1.08/1337\tablenotemark{c}
\enddata

\tablenotetext{a}{Absorbed flux in 0.5-7\,keV range.}
\tablenotetext{b}{Unabsorbed flux ratio between the power-law and the
primary (i.e., blackbody, STEMS, or RCS) components in 0.5-7\,keV range.}
\tablenotetext{c}{All nine data sets were fitted jointly with a single column
density.}
\tablenotetext{d}{Held fixed in the fit.}
\tablecomments{All uncertainties quoted are 90\% confidence intervals
(i.e., 1.6$\sigma$).}
\end{deluxetable*}

To look for energy dependence of the modulation, we
estimated the PFs in the soft (1-3\,keV) and hard (3-7\,keV)
energy bands separately. In the 2008 data, the latter shows a
systematically higher PF, with a difference ranging from
$\Delta$PF = 0.04-0.09 (i.e., a 20\%-30\% change), which is
statistically significant given the measurement uncertainty is
only $\sim$0.01. However, such an energy dependence is not observed
in 2009, with the PFs in the two bands being consistent with each other.

\subsubsection{{\it Chandra} Spectroscopy}

The \cxo\ spectra of \axp\ were extracted using the tool
\texttt{psextract} in CIAO, then binned such that every bin has an
S/N of at least 10. We performed the spectral fits in the 0.5-7\,keV
range with XSPEC v12.6.0. All nine datasets were fitted jointly
with a single absorption column density ($N_{\rm H}$). We 
also tried fitting different $N_{\rm H}$ values for the 2008 and 2009
data, and confirmed that they are consistent. We started with
simple models including an absorbed blackbody (BB) and an
absorbed power-law (PL), but obtained very poor fits (reduced
$\chi^2$ values over 1.5). An absorbed blackbody plus power-law
(BB+PL) model gives much better fits and the results are listed in
Table~\ref{t2} and plotted in Figure~\ref{f3}. As shown in the
figure, the fit residuals suggest a hint of a spectral feature
$\sim1.3$\,keV, which is more obvious in 2009 than in 2008.
However, the significance is only $\sim$1$\sigma$ and deeper
exposures are needed to confirm this.

In addition to the BB+PL model, we also considered more physical
models that account for the Compton up-scattering of the thermal
photons in the magnetosphere. We tried fitting the Resonant
Cyclotron Scattering \citep[RCS;][]{lg06,rzt+08} and the Surface Thermal
Emission and Magnetospheric Scattering \citep[STEMS;][]{o03,gol06}
models to the data. In the latter, the gravitational redshift is
fixed at $z=0.306$ during the fit, corresponding to the canonical
neutron star mass of 1.4\,$M_\odot$ and radius of 10\,km. While
these models fit the 2008 data reasonably well, the 2009 spectra
clearly require an additional hard component. Therefore, we added
a PL to the spectral model in 2009, with $\Gamma$ fixed at 1.33
according to the \emph{Suzaku} results \citep{enm+10}.\footnote{The
\emph{INTEGRAL} results also suggest that the hard-band PL
spectral index remained stable over the period of the \cxo\ 
observations \citep{dkh09}.} Compared to the BB+PL fit above,
these models provide a similar goodness-of-fit in terms of the
reduced $\chi^2$ values. Table~\ref{t2} lists the key parameters
of the best-fit models. The scattering optical depth $\tau$ is
around 1-2 in 2008 and $\gtrsim3$ in 2009, and the thermal
velocity $\beta$ of the electrons is $\sim$0.4-0.5 in 2008 and
$\sim0.2$ in 2009 \citep[see][for a detailed definition of these
parameters]{lg06}.

\section{Discussion}
\label{s3} 
In this paper, we have reported on \cxo\ observations of \axp\
immediately following its 2008 and 2009 outbursts, along with
{\it RXTE} timing and pulsed flux behavior following the 2008
outburst and throughout the 2009 event. Next we discuss these
observations in the context of the magnetar model.

\subsection{Spectral and Spin Evolution}

In the twisted magnetosphere model of magnetars \citep{tlk02},
the observed X-ray luminosity of a magnetar is determined both
by its surface temperature and by magnetospheric currents, the
latter due to the twisted dipolar field structure. The surface
temperature in turn is determined by the energy output from
within the star due to magnetic field decay, as well as on the
nature of the atmosphere and the stellar magnetic field strength.
This surface thermal emission is resonantly scattered by the
current particles, thus resulting in an overall spectrum similar
to a Comptonized blackbody \citep[e.g.][]{lg06,rzt+08,zrt+09}.
In this model, the greater the twist angle, the greater the
scattering, the harder the spectrum, and the greater the X-ray
luminosity $L_X$. In addition, the surface heating by return
currents is believed to contribute substantially to $L_X$,
at least at the same level as the thermal component induced
from the interior field decay \citep{tlk02}. Magnetar outbursts in this
picture occur with sudden increases in twist angle, consistent
with the generic hardening of magnetar spectra during outbursts
\citep[e.g.][]{kgw+03,wkt+04,icd+07}.

Other observational evidence provided in support of the
twisted magnetosphere model as proposed by \citet{tlk02}
is a correlation between magnetar spectral hardness and
spin-down-inferred magnetic field strength $B$, when comparing
different sources \citep{mw01,kb10}. In this case, $B$ is an
observational proxy for the magnetospheric twist angle. On the
other hand, some magnetars have shown dramatic spin-down rate
variations, with order-of-magnitude changes in $\dot{\nu}$ 
seen on a variety of timescales \cite[e.g.][]{gk04,wkt+04}.
The origin of these variations is unknown. Nevertheless, in
the context of the twisted magnetosphere model, a varying twist
angle might naively be expected to be accompanied by a
changing $\dot{\nu}$ (due to changing effective $B$), and
corresponding spectral and flux changes. However, some
decoupling between $\dot{\nu}$ and the radiative behavior
might be expected, particularly as the spin-down is affected most
by a narrow field-line bundle near the light cylinder, whereas the
radiation originates from the surface. Field-line twists likely
propagate outward \citep{tlk02}, which suggests the radiative changes
should occur prior to $\dot{\nu}$ changes \citep{bt07}.

In contrast to the picture in which a magnetar outburst is
accompanied by an enhanced magnetospheric twist, \citet{og07}
suggest that in outburst, the magnetosphere may be stable,
with radiative evolution being due to changes in the surface
thermal emission. Using a spectral model consisting of a
resonant Comptonized atmosphere-modified blackbody (the STEMS model), 
fits to data for XTE J1810$-$197 \citep{gog+07}
result in the spectrally inferred $B$ being stable, with all
radiative changes being due to changes in the surface thermal emission. 

For \axp, the source spectrum only showed significant variability
over a short period of time ($\sim$1\,day) following the 2008 and
2009 outbursts (\citealt{ier+10}; Scholz et al., in preparation),
but remained stable over the \cxo\ observations, during which the
flux changed substantially (see Table~\ref{t2}). The latter seems
opposite to the hardness/flux correlation predicted by the twisted
magnetosphere model. On the other hand, we note that the flux decay
over the \cxo\ epochs has a comparable timescale to that of the pulse
profile variations (Figure~\ref{f2}), which generally agrees with
the predictions of the magnetar model. More intriguing are the timing
results. As described in Section~\ref{sec:rxtetiming}, we found a factor
of 2.2 increase in the spin-down rate $|\dot{\nu}|$ between the first
and last \cxo\ observations in 2008, a substantial change even by
magnetar standards. For purely dipole
spin-down, this naively implies a $\sim$50\% increase in the
effective surface dipole field strength, from an initial
spin-inferred value of $2.8\times 10^{14}$\,G at the epoch of the
first {\it Chandra} observation, to a value of $4.1\times10^{14}$\,G
at the epoch of the last. This is contrary to SGR~1806$-$20, in
which the spectral response lagged behind the torque variation,
suggesting some hysteresis in the system \citep{wkf+07}. In our
case, the lack of associated spectral changes is unexpected in
the twisted magnetosphere model, unless the spin-down is
decoupled from the site of the radiative events, as suggested
by \citet{gk04} for 1E 1048.1$-$5937 \citep[see also][]{b09}.
Although $B$ fields inferred from long-term spin-down have been
used to compare with those inferred spectrally \citep[e.g.][]{gog+07},
our results call into question the reliability of such a comparison
when using short-term spin-down rate.

Indeed, for the 2008 \cxo\ observations of \axp, spectral fits
using the STEMS model yield a very similar $B$ field value for
all the observations (see Table~\ref{t2}) in spite of the strongly
varying $\dot{\nu}$. One way to reconcile this is to interpret the
$B$ field measured spectroscopically as higher-order multipoles in
localized X-ray-emitting regions rather than the global dipole
field responsible for spin-down. However, we note that the $B$ field
obtained from the STEMS model is \emph{lower} than the
spin-down-inferred value at the epoch of the last 2008 \cxo\
observation, which is difficult to be explained by the above picture.
As an alternative, the extra spin-down torque could
be attributed to particle winds \citep{hck99}. Comparing the
spin-down rates of \axp\ in 2007 \citep{crj+08}, 2008 and 2009
(Section~\ref{sec:rxtetiming}), it is obvious that the spin-down torque
changed drastically between these epochs. Torque variations have
been observed in magnetars and ordinary radio pulsars
\citep[e.g.][]{kgw+03,gk04,klo+06,lhk+10,lnk+11}.
These could be related to changes in plasma conditions in the
magnetosphere, which may not necessarily have any observable
effects in the radiative properties \citep[see][]{lnk+10}. Based
on Equation~(9) in \citet{hck99}, a factor of 2.2 increase in
$|\dot{\nu}|$, and hence in $\dot{E}$, requires a steady wind
luminosity ($L_p$) 1.5 times larger than the dipole spin-down
luminosity, implying $L_p\approx1.5\times10^{35}\,$\ergs. For a
typical X-ray efficiency of below 1\%, the particle-induced flux
would be less than $10^{-12}$\erg, much smaller when compared
to the source's flux (see Table~\ref{t2}), thus, unlikely to be
detected.

In any case, the absence of spectral variations in the presence
of flux changes remains puzzling whether or not the magnetospheric
twist angle varied in the outburst. Moreover, a reasonable
model will need to explain why $\dot\nu$ changed drastically 
following the 2008 event, but stayed constant in 2009. This
may reinforce the requirement of decoupling between the spin-down
and the source of the radiative changes, hence presumably
between the spin-down-inferred magnetic field and that inferred spectrally.

\begin{figure}[ht]
\epsscale{1.0}
\plotone{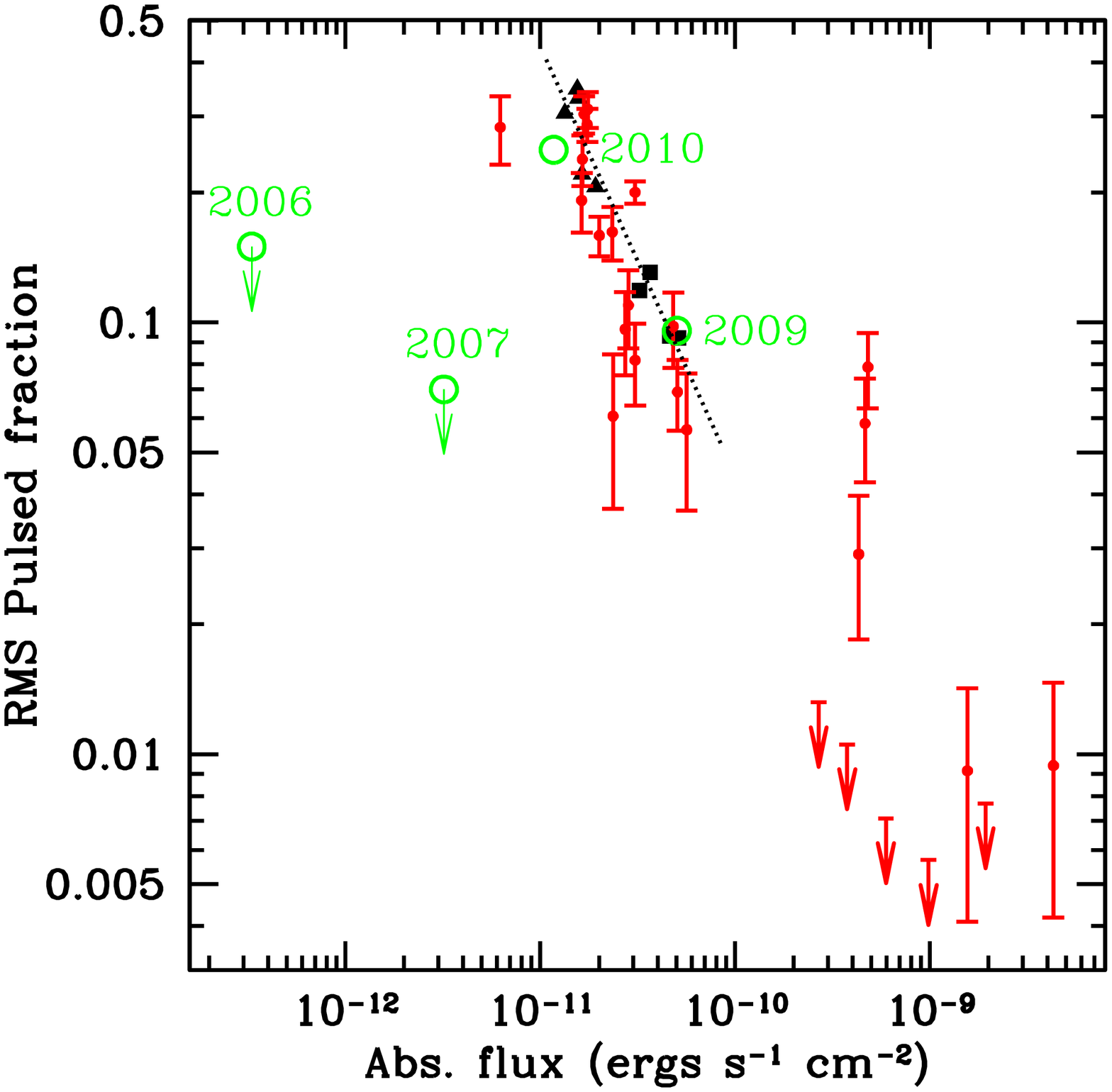}
\caption{rms PF vs.\ absorbed flux of \axp\ in
the 1-7\,keV band.  The filled triangles and squares show the 2008 and
2009 \cxo\ observations, respectively. The green open circle and
red dots represent results from \xmm\ and \emph{Swift}, respectively
(\citealt{hgr+08}; S.\ A.\ Olausen et al., in preparation; P.\ Scholz
et al., in preparation), with the \xmm\ epochs labeled. Note that
the 2007 and 2008 \xmm\ data points are only upper limits, since
the values reported in the literature are area PF estimates, which
are larger than the rms PFs by a constant which depends on the pulse
shape. The dashed line shows a power-law fit to the \cxo\ data.
\label{f4}}
\end{figure}

\subsection{Flux Evolution During the 2008 and 2009 Events}
The 2008 and 2009 events exhibited very different flux evolutions.
Immediately after the 2009 outburst, the persistent flux
increased by a factor of $\sim500$ (Scholz et al.\ in preparation),
while the pulsed flux evidently showed only very little
variation (less than a factor of two). We found a monotonic
flux decay during the 2009 recovery, with a power-law of
index $-0.21\pm0.01$, which is comparable to $-0.306\pm0.005$ for
CXOU J164710.2$-$455216 \citep{wkg+11}, but not as steep as
$-0.69\pm0.03$ for 1E~2259+586 \citep{zkd+08} or $-0.92\pm0.02$
for XTE J1810$-$197 \citep{wkg+05}.

In contrast, the 2008 event is less energetic; the total flux
increased only by a factor of $\sim$100 \citep{ier+10}, but our
\emph{RXTE} results reveal a pulsed flux variation by a factor
of $\sim$4, far greater than in the 2009 event. Also, the flux
decay in 2008 showed a more complicated history. As is clear
in Figure~\ref{f1}, we observed an additional flux enhancement
around MJD 54752, $\sim$11 days after the initial trigger, 
\footnote{Although this second enhancement is not reported by
\citet{ier+10} in their study of the {\it Swift} observations
covering the same period, we have re-analysed the same {\it Swift}
dataset and confirmed the pulsed flux enhancement we observe with
{\it RXTE} and {\it Chandra} (P.\ Scholz et al., in preparation).}
lasting for $\sim$30 days. 

\citet{es10} suggest that radiative outbursts in magnetars could
generally be preceded by glitches, with the delay between the
two events depending on the depth at which the glitch-induced
energy release occurs. For the initial event in 2008, we are
unable to tell whether a glitch preceded the radiative outburst,
due to the lack of {\it RXTE} observations prior to the
outburst. However, we can rule out any glitch between the
initial 2008 event and the second flux enhancement 10 days
later. It is possible that the initial event actually involved
glitches occurring in two different places in the stellar interior,
at substantially different depths, such that the delays from the
glitch to the X-ray enhancement were different, but we note
that this picture does not explain the sharp rise of the
second flux enhancement.

\subsection{Pulsed Fraction Evolution}
Our results in Section~\ref{s2} clearly indicate a strong
anti-correlation between the PF and the phase-averaged X-ray flux, at
least during the 2008 and 2009 outbursts. This is plotted in
Figure~\ref{f4}, and suggests an approximate power-law relation between
the two observables. The trend is also supported by the \xmm\ and
\emph{Swift} measurements taken in the same period (S.\ A.\ Olausen et
al., in preparation, P.\ Scholz et al., in preparation).
Similar anti-correlations have
been observed in 1E 1048.1$-$5937 and CXOU J164710.2$-$455216
\citep{tgd+08,icd+07}, while positive correlations were found in
XTE J1810$-$197 and 1E~2259+586 \citep{ghb+04,zkd+08}. This variety
of behaviors is consistent with the picture in which, from source
to source, the location and geometry of the region on the star
affected in the outburst are different. Previous studies proposed
that an anti-correlation could be the consequence of an increased
emitting area due to an outburst, such that part of the hot spot
becomes visible at any phase, thus reducing the pulse modulation
\citep[e.g.][]{hgr+08}. We note that this scenario depends critically
on the location of emission zone on the stellar surface as well as
the viewing geometry \citep[see][]{bgr08}; it may be possible
to obtain either a monotonic increase or decrease of PF
for the same area `hot spot' depending on its location
on the stellar surface.

We point out that the trend of decreasing PF for
increasing phase-averaged X-ray flux observed in the \cxo\ data
does not seem to hold for other observations of \axp\ before 2008.
Based on \xmm\ exposures, \citet{hgr+08} reported area PFs of
15\% in quiescence in 2006 and 7\% in the high state in 2007. As
shown in Figure~\ref{f4}, these values deviate significantly from
the 2008-2009 trend. The discrepancy seems too large to be reconciled
by a difference in the instrument response, and is even larger if the
differing PF estimate methods are accounted for. It is possible that
the outburst in 2008 induced some permanent changes in the $B$-field
configuration or emission geometry. The pulse profiles shown in
\citet{hgr+08} also appear to have a different shape from the ones
shown in Figure~\ref{f2}, providing further support to this picture.

\section{Conclusions}
\label{s4} 
We have presented results from \cxo\ and \emph{RXTE} observations of
\axp\ following its 2008 and 2009 outbursts. These allow a
direct comparison between the two events. We found that over
the 2008 \cxo\ observation epochs, the pulsar spin-down rate
increased by a factor 2.2, in the absence of corresponding
spectral changes, whereas such variation in $\dot\nu$ is not
observed after the more energetic 2009 event. This provides
evidence of decoupling between magnetar spin and radiative
properties.  
The absence of spectral changes simultaneous with significant
flux decay is surprising for models of magnetar outbursts.
Our results also revealed a strong anti-correlation
between the PF and phase-averaged flux of the source.
While both 2008 and 2009 data follow the same trend, pre-2008
measurements show significant deviation,
suggesting that the 2008 outburst may have induced permanent
changes in the emission geometry. Finally, we note that \axp\ 
demonstrated significant spectral changes only
within the first day after the 2008 and 2009 events, which
highlights the importance of prompt observations in future
studies for understanding post-outburst relaxation of magnetars.  

\acknowledgements
We thank A. Archibald, S.\ Bogdanov, M.\ Livingstone and W.\ Zhu for useful
discussions. C.-Y.N.\ is a CRAQ postdoctoral fellow.
V.M.K.\ holds the Lorne Trottier Chair in Astrophysics and Cosmology
and a Canada Research Chair in Observational Astrophysics.
This work is supported by an NSERC Discovery Grant, by CIFAR, and by
FQRNT via CRAQ.

{\it Facilities:}
\facility{CXO (ACIS)}
\facility{RXTE (PCA)}


\end{document}

%% file: def.tex
\newcommand{\cxo}{\emph{Chandra}}
\newcommand{\xmm}{\emph{XMM-Newton}}

\newcommand{\ergs}{\,ergs\,s$^{-1}$}
\newcommand{\erg}{\,ergs\,cm$^{-2}$\,s$^{-1}$}

%% file: arxiv.bbl
\begin{thebibliography}{49}
\expandafter\ifx\csname natexlab\endcsname\relax\def\natexlab#1{#1}\fi

\bibitem[{{Beloborodov}(2009)}]{b09}
{Beloborodov}, A.~M. 2009, \apj, 703, 1044

\bibitem[{{Beloborodov} \& {Thompson}(2007)}]{bt07}
{Beloborodov}, A.~M., \& {Thompson}, C. 2007, \apj, 657, 967

\bibitem[{{Bogdanov} {et~al.}(2008){Bogdanov}, {Grindlay}, \&
  {Rybicki}}]{bgr08}
{Bogdanov}, S., {Grindlay}, J.~E., \& {Rybicki}, G.~B. 2008, \apj, 689, 407

\bibitem[{{Camilo} {et~al.}(2007){Camilo}, {Ransom}, {Halpern}, \&
  {Reynolds}}]{crh+07}
{Camilo}, F., {Ransom}, S.~M., {Halpern}, J.~P., \& {Reynolds}, J. 2007, \apjl,
  666, L93

\bibitem[{{Camilo} {et~al.}(2008){Camilo}, {Reynolds}, {Johnston}, {Halpern},
  \& {Ransom}}]{crj+08}
{Camilo}, F., {Reynolds}, J., {Johnston}, S., {Halpern}, J.~P., \& {Ransom},
  S.~M. 2008, \apj, 679, 681

\bibitem[{{den Hartog} {et~al.}(2009){den Hartog}, {Kuiper}, \&
  {Hermsen}}]{dkh09}
{den Hartog}, P.~R., {Kuiper}, L., \& {Hermsen}, W. 2009, ATel, 1922

\bibitem[{{Dib} {et~al.}(2008){Dib}, {Kaspi}, \& {Gavriil}}]{dkg08}
{Dib}, R., {Kaspi}, V.~M., \& {Gavriil}, F.~P. 2008, in AIP Conf.\ Proc.\ 
983, 40 Years of Pulsars: Millisecond
  Pulsars, Magnetars and More, ed. {C.~Bassa, Z.~Wang, A.~Cumming, \&
  V.~M.~Kaspi}, 239

\bibitem[{{Dib} {et~al.}(2009){Dib}, {Kaspi}, \& {Gavriil}}]{dkg09}
{Dib}, R., {Kaspi}, V.~M., \& {Gavriil}, F.~P. 2009, \apj, 702, 614

\bibitem[{{Eichler} \& {Shaisultanov}(2010)}]{es10}
{Eichler}, D., \& {Shaisultanov}, R. 2010, \apjl, 715, L142

\bibitem[{{Enoto} {et~al.}(2010){Enoto}, {Nakazawa}, {Makishima}, {Nakagawa},
  {Sakamoto}, {Ohno}, {Takahashi}, {Terada}, {Yamaoka}, {Murakami}, \&
  {Takahashi}}]{enm+10}
{Enoto}, T., {et~al.} 2010, \pasj, 62, 475

\bibitem[{{Gavriil} \& {Kaspi}(2004)}]{gk04}
{Gavriil}, F.~P., \& {Kaspi}, V.~M. 2004, \apjl, 609, L67

\bibitem[{{Gavriil} {et~al.}(2006){Gavriil}, {Kaspi}, \& {Woods}}]{gkw06}
{Gavriil}, F.~P., {Kaspi}, V.~M., \& {Woods}, P.~M. 2006, \apj, 641, 418

\bibitem[{{Gelfand} \& {Gaensler}(2007)}]{gg07}
{Gelfand}, J.~D., \& {Gaensler}, B.~M. 2007, \apj, 667, 1111

\bibitem[{{Gotthelf} {et~al.}(2004){Gotthelf}, {Halpern}, {Buxton}, \&
  {Bailyn}}]{ghb+04}
{Gotthelf}, E.~V., {Halpern}, J.~P., {Buxton}, M., \& {Bailyn}, C. 2004, \apj,
  605, 368

\bibitem[{{Gregory} \& {Loredo}(1992)}]{gl92}
{Gregory}, P.~C., \& {Loredo}, T.~J. 1992, \apj, 398, 146

\bibitem[{{Gronwall} {et~al.}(2009){Gronwall}, {Holland}, {Markwardt},
  {Palmer}, {Stamatikos}, \& {Vetere}}]{ghm+09}
{Gronwall}, C., {Holland}, S.~T., {Markwardt}, C.~B., {Palmer}, D.~M.,
  {Stamatikos}, M., \& {Vetere}, L. 2009, GRB Coord. Netw., 8833, 1

\bibitem[{{G{\"u}ver} {et~al.}(2007){G{\"u}ver}, {{\"O}zel}, {G{\"o}{\u
  g}{\"u}{\c s}}, \& {Kouveliotou}}]{gog+07}
{G{\"u}ver}, T., {{\"O}zel}, F., {G{\"o}{\u g}{\"u}{\c s}}, E., \&
  {Kouveliotou}, C. 2007, \apjl, 667, L73

\bibitem[{{G\"uver} {et~al.}(2006){G\"uver}, {\"Ozel}, \& {Lyutikov}}]{gol06}
{G\"uver}, T., {\"Ozel}, F., \& {Lyutikov}, M. 2006, arXiv:astro-ph/0611405

\bibitem[{{Halpern} {et~al.}(2008){Halpern}, {Gotthelf}, {Reynolds}, {Ransom},
  \& {Camilo}}]{hgr+08}
{Halpern}, J.~P., {Gotthelf}, E.~V., {Reynolds}, J., {Ransom}, S.~M., \&
  {Camilo}, F. 2008, \apj, 676, 1178

\bibitem[{{Harding} {et~al.}(1999){Harding}, {Contopoulos}, \&
  {Kazanas}}]{hck99}
{Harding}, A.~K., {Contopoulos}, I., \& {Kazanas}, D. 1999, \apjl, 525, L125

\bibitem[{{Israel} {et~al.}(2007){Israel}, {Campana}, {Dall'Osso}, {Muno},
  {Cummings}, {Perna}, \& {Stella}}]{icd+07}
{Israel}, G.~L., {Campana}, S., {Dall'Osso}, S., {Muno}, M.~P., {Cummings}, J.,
  {Perna}, R., \& {Stella}, L. 2007, \apj, 664, 448

\bibitem[{{Israel} {et~al.}(2010){Israel}, {Esposito}, {Rea}, {Dall'Osso},
  {Senziani}, {Romano}, {Mangano}, {G{\"o}tz}, {Zane}, {Tiengo}, {Palmer},
  {Krimm}, {Gehrels}, {Mereghetti}, {Stella}, {Turolla}, {Campana}, {Perna},
  {Angelini}, \& {De Luca}}]{ier+10}
{Israel}, G.~L., {et~al.} 2010, \mnras, 408, 1387

\bibitem[{{Kaneko} {et~al.}(2010){Kaneko}, {G{\"o}{\u g}{\"u}{\c s}},
  {Kouveliotou}, {Granot}, {Ramirez-Ruiz}, {van der Horst}, {Watts}, {Finger},
  {Gehrels}, {Pe'er}, {van der Klis}, {von Kienlin}, {Wachter}, {Wilson-Hodge},
  \& {Woods}}]{kgk+10}
{Kaneko}, Y., {et~al.} 2010, \apj, 710, 1335

\bibitem[{{Kaspi}(2007)}]{kas07}
{Kaspi}, V.~M. 2007, \apss, 308, 1

\bibitem[{{Kaspi} \& {Boydstun}(2010)}]{kb10}
{Kaspi}, V.~M., \& {Boydstun}, K. 2010, \apjl, 710, L115

\bibitem[{{Kaspi} {et~al.}(2003){Kaspi}, {Gavriil}, {Woods}, {Jensen},
  {Roberts}, \& {Chakrabarty}}]{kgw+03}
{Kaspi}, V.~M., {Gavriil}, F.~P., {Woods}, P.~M., {Jensen}, J.~B., {Roberts},
  M.~S.~E., \& {Chakrabarty}, D. 2003, \apjl, 588, L93

\bibitem[{{Kramer} {et~al.}(2006){Kramer}, {Lyne}, {O'Brien}, {Jordan}, \&
  {Lorimer}}]{klo+06}
{Kramer}, M., {Lyne}, A.~G., {O'Brien}, J.~T., {Jordan}, C.~A., \& {Lorimer},
  D.~R. 2006, Science, 312, 549

\bibitem[{{Kuiper} {et~al.}(2009){Kuiper}, {den Hartog}, \& {Hermsen}}]{kdh09}
{Kuiper}, L., {den Hartog}, P.~R., \& {Hermsen}, W. 2009, ATel, 1921

\bibitem[{{Lamb} \& {Markert}(1981)}]{lm81}
{Lamb}, R.~C., \& {Markert}, T.~H. 1981, \apj, 244, 94

\bibitem[{{Livingstone} {et~al.}(2011){Livingstone}, {Ng}, {Kaspi}, {Gariil},
  \& {Gotthelf}}]{lnk+11}
{Livingstone}, L., {Ng}, C.-Y., {Kaspi}, V.~M., {Gariil}, F.~P., \& {Gotthelf},
  E.~V. 2011, \apj, submitted (arXiv:1007.2829)

\bibitem[{{Lyne} {et~al.}(2010){Lyne}, {Hobbs}, {Kramer}, {Stairs}, \&
  {Stappers}}]{lhk+10}
{Lyne}, A., {Hobbs}, G., {Kramer}, M., {Stairs}, I., \& {Stappers}, B. 2010,
  Science, 329, 408

\bibitem[{{Lyutikov} \& {Gavriil}(2006)}]{lg06}
{Lyutikov}, M., \& {Gavriil}, F.~P. 2006, \mnras, 368, 690

\bibitem[{{Marsden} \& {White}(2001)}]{mw01}
{Marsden}, D., \& {White}, N.~E. 2001, \apjl, 551, L155

\bibitem[{{Mereghetti}(2008)}]{mer08}
{Mereghetti}, S. 2008, A\&AR, 15, 225

\bibitem[{{Mereghetti} {et~al.}(2009){Mereghetti}, {G{\"o}tz},
  {Weidenspointner}, {von Kienlin}, {Esposito}, {Tiengo}, {Vianello}, {Israel},
  {Stella}, {Turolla}, {Rea}, \& {Zane}}]{mgw+09}
{Mereghetti}, S., {et~al.} 2009, \apjl, 696, L74

\bibitem[{{{\"O}zel}(2003)}]{o03}
{{\"O}zel}, F. 2003, \apj, 583, 402

\bibitem[{{{\"O}zel} \& {G{\"u}ver}(2007)}]{og07}
{{\"O}zel}, F., \& {G{\"u}ver}, T. 2007, \apjl, 659, L141

\bibitem[{{Rea} {et~al.}(2008){Rea}, {Zane}, {Turolla}, {Lyutikov}, \&
  {G{\"o}tz}}]{rzt+08}
{Rea}, N., {Zane}, S., {Turolla}, R., {Lyutikov}, M., \& {G{\"o}tz}, D. 2008,
  \apj, 686, 1245

\bibitem[{{Savchenko} {et~al.}(2010){Savchenko}, {Neronov}, {Beckmann},
  {Produit}, \& {Walter}}]{snb+10}
{Savchenko}, V., {Neronov}, A., {Beckmann}, V., {Produit}, N., \& {Walter}, R.
  2010, \aap, 510, A77

\bibitem[{{Tam} {et~al.}(2008){Tam}, {Gavriil}, {Dib}, {Kaspi}, {Woods}, \&
  {Bassa}}]{tgd+08}
{Tam}, C.~R., {Gavriil}, F.~P., {Dib}, R., {Kaspi}, V.~M., {Woods}, P.~M., \&
  {Bassa}, C. 2008, \apj, 677, 503

\bibitem[{{Terada} {et~al.}(2009){Terada}, {Tashiro}, {Urata}, {Endo}, {Onda},
  {Kodaka}, {Morigami}, {Sugasahara}, {Iwakiri}, {Ohno}, {Kokubun}, {Suzuki},
  {Takahashi}, {Yamaoka}, {Sugita}, {Enoto}, {Nakazawa}, {Makishima},
  {Nakagawa}, {Tamagawa}, {Uehara}, {Fukazawa}, {Kira}, {Hanabata}, {Sonoda},
  {Yamauchi}, {Tanaka}, {Hara}, {Ohmori}, {Kono}, {Hayashi}, {Hong}, \&
  {Vasquez}}]{ttu+09}
{Terada}, Y., {et~al.} 2009, GRB Coord. Netw., 8845, 1

\bibitem[{{Thompson} {et~al.}(2002){Thompson}, {Lyutikov}, \&
  {Kulkarni}}]{tlk02}
{Thompson}, C., {Lyutikov}, M., \& {Kulkarni}, S.~R. 2002, \apj, 574, 332

\bibitem[{{Tiengo} {et~al.}(2010){Tiengo}, {Vianello}, {Esposito},
  {Mereghetti}, {Giuliani}, {Costantini}, {Israel}, {Stella}, {Turolla},
  {Zane}, {Rea}, {G{\"o}tz}, {Bernardini}, {Moretti}, {Romano}, {Ehle}, \&
  {Gehrels}}]{tve+10}
{Tiengo}, A., {et~al.} 2010, \apj, 710, 227

\bibitem[{{Woods} {et~al.}(2011){Woods}, {Kaspi}, {Gavriil}, \&
  {Airhart}}]{wkg+11}
{Woods}, P.~M., {Kaspi}, V.~M., {Gavriil}, F.~P., \& {Airhart}, C.\ 2011, \apj,
  submitted (arXiv:1006.2950)

\bibitem[{{Woods} {et~al.}(2007){Woods}, {Kouveliotou}, {Finger}, {G{\"o}{\u
  g}{\"u}{\c s}}, {Wilson}, {Patel}, {Hurley}, \& {Swank}}]{wkf+07}
{Woods}, P.~M., {Kouveliotou}, C., {Finger}, M.~H., {G{\"o}{\u g}{\"u}{\c s}},
  E., {Wilson}, C.~A., {Patel}, S.~K., {Hurley}, K., \& {Swank}, J.~H. 2007,
  \apj, 654, 470

\bibitem[{{Woods} {et~al.}(2004){Woods}, {Kaspi}, {Thompson}, {Gavriil},
  {Marshall}, {Chakrabarty}, {Flanagan}, {Heyl}, \& {Hernquist}}]{wkt+04}
{Woods}, P.~M., {et~al.} 2004, \apj, 605, 378

\bibitem[{{Woods} {et~al.}(2005){Woods}, {Kouveliotou}, {Gavriil}, {Kaspi},
  {Roberts}, {Ibrahim}, {Markwardt}, {Swank}, \& {Finger}}]{wkg+05}
---. 2005, \apj, 629, 985

\bibitem[{{Zane} {et~al.}(2009){Zane}, {Rea}, {Turolla}, \& {Nobili}}]{zrt+09}
{Zane}, S., {Rea}, N., {Turolla}, R., \& {Nobili}, L. 2009, \mnras, 398, 1403

\bibitem[{{Zhu} {et~al.}(2008){Zhu}, {Kaspi}, {Dib}, {Woods}, {Gavriil}, \&
  {Archibald}}]{zkd+08}
{Zhu}, W., {Kaspi}, V.~M., {Dib}, R., {Woods}, P.~M., {Gavriil}, F.~P., \&
  {Archibald}, A.~M. 2008, \apj, 686, 520

\end{thebibliography}
